\documentstyle[12pt,amsfonts]{article}
\newcommand{\svec}[1]{ \stackrel{\rightarrow}{#1} }

\def\be{\begin{equation}}
\def\ee{\end{equation}}
\def\ba{\begin{array}}
\def\ea{\end{array}}

\def\d4{{\rm d}^4}

\begin{document}
%---------------------------------------------------
\title{\bf Determination of Gravitomagnetic Field through GRBs
            or X-ray Pulsars}
\author{ {Ning WU}\thanks{email address: wuning@mail.ihep.ac.cn}
{, Dahua ZHANG}
\\
\\
{\small Institute of High Energy Physics, P.O.Box 918-1,
Beijing 100039, P.R.China}
}
\maketitle
%\vskip 0.8in
%\noindent

%\vskip 0.8in
%\noindent

\begin{abstract}
In gauge theory of gravity, there is direct coupling
between the spin of a particle and gravitomagnetic field.
In the surface of a neutron star or near black hole,
the coupling energy between spin and gravitomagnetic field
can be large and detectable. Precise measurement of the
position of spectrum lines of the corresponding emission or
absorption can help us to determine the gravitomagnetic field and
electromagnetic field simultaneously. The ratio
$\frac{\Delta E_e}{\Delta E_p}$
 can be served as a quantitative criteria of black hole.
In GRBs or X-ray pulsar, absorption spectral
lines of electron were observed. If the absorption spectral lines
of electron and proton can be observed simultaneously, using
the method given in this paper, we can determine the gravitomagnetic
field in the surface of the star, and discriminate black hole
from neutron star.
\\

\end{abstract}

~~

%-------------------------------------------------------

\newpage

\Roman{section}

\section{Introduction}

The classical effects of gravitomagnetism were studied for more
than one hundred years.
The close analogy between Newton's gravitation law and Coulomb's law
of electricity led to the birth of the concept of gravitomagnetism
in the nineteenth century\cite{m01,m02,m03,m04}. Later, in the
framework of General Relativity, gravitomagnetism was extensively
explored\cite{m05,m06,m07} and recently some experiments are
designed to test gravitomagnetism effects. Some recently reviews
on gravitomagnetism can be found in literatures \cite{m08,m09,m10}.
\\

In quantum gauge theory of gravity\cite{6,7,8,9,10,11},
gravitoelectric field and gravitomagnetic field are naturally
defined as component of field strength of gravitational gauge field.
In gauge theory of gravity, the definition of gravitomagnetism
is more general, that is, the gravitomagnetism defined in
the literature \cite{m05,m06,m07,m08,m09,m10} only corresponds
to the time components of the gravitomagnetism defined in
gauge theory of gravity. It is known that gravitational Lorentz
force is a direct classical gravitomagnetic effect
of a moving mass point in gravitomagnetic field\cite{12,13}.
\\

In classical level, gauge theory of gravity returns to
general relativity. In quantum level,
gauge theory of gravity is a perturbatively renormalizable
quantum theory, so based on it, quantum effects of gravity
can be explored. Based on a unified theory of gravity and
electromagnetic interactions\cite{14,15,16}, some known
quantum effects related to gravitational interactions are
discussed in the literature \cite{17}. In this paper, emission
and absorption of a spin particle in strong gravitomagnetic
field is studied. A proposal to detect such kind of
emission or absorption is given and the meaning of such
kind of measurement of gravitomagnetism is discussed.
Determining the gravitomagnetic
field of a star  can help us to known which star
is black hole, for can only black hole  have very strong
gravitomagnetic field.
\\

\section{Coupling between spin and gravitomagnetic field}
\setcounter{equation}{0}

In gauge theory of  gravity, gravitational field is represented
by gravitational gauge field $C_{\mu}^{\alpha}$. Its field
strength is $F_{\mu\nu}^{\alpha}$. The gravitational field
and gravitoelectric field are defined by
\be \label{2.1}
E_i = F_{0i} = E_i^{\alpha} \hat P_{\alpha},
\ee
\be \label{2.2}
B_i = - \frac{1}{2} \varepsilon_{ijk} F_{jk} =
 B_i^{\alpha} \hat P_{\alpha},
\ee
where $\hat P_{\alpha}$ is the generator of global gravitational
gauge group.
\\

The equation of motion of a Dirac particle in both electromagnetic
field and gravitational field is\cite{17}
\be \label{2.201}
\left\lbrack \gamma^{\mu} ( \partial_{\mu} - ie A_{\mu}
- g C_{\mu}^{\alpha} \partial_{\alpha} ) + m \right\rbrack
\psi = 0.
\ee
Making non-relativistic limit of the above equation, we can obtain
the following
Schrodinger equation for a non-relativistic Dirac particle
in strong gravitomagnetic field and electromagnetic magnetic
field is\cite{17}
\be \label{2.3}
\ba{rcl}
i \frac{\partial \Psi}{\partial t} &=& \left\lbrack
\frac{1}{2m} \left (-i \svec{D} - e \svec{A}
        + m g \svec{C}^0 \right)^2
+ m g C_0^0 - e A_0  \right.\\
&&\\
&& \left.
- \frac{g}{2} \svec{\sigma} \cdot \svec{B}^0
- \frac{e}{2 m} \svec{\sigma} \cdot \svec{B}_e
- \frac{i g}{2}  \svec{\sigma} \cdot \svec{E}^0
- \frac{i e }{ 2 m}\svec{\sigma} \cdot \svec{E}_e
 \right\rbrack \Psi,
\ea
\ee
where
\be \label{2.4}
\svec{C}^0 = (C_i^0),
\ee
$\svec{B}_e$ and  $\svec{E}_e$ are electromagnetic field,
$\svec{\sigma}$ is the spin matrix of the Dirac particle, $g$ and $e$
the coupling constant of gravitational interactions and electromagnetic
interactions respectively,
$A_{\mu}$ is electromagnetic gauge  field,
and $\svec{A}$ is its vector potential.
It could be seen that the classical Newtonian gravitational potential
naturally enters into the Schrodinger equation. Besides, there is direct
coupling between spin and gravitomagnetic field, no matter the
Dirac particle carries electric charge or not. \\

When there are strong gravitomagnetic field and electromagnetic
magnetic field, there is coupling  between spin and gravitomagnetic
field and electromagnetic magnetic field, which has the following
coupling energy
\be \label{2.5}
\Delta H =
 - \frac{g}{2} \svec{\sigma} \cdot \svec{B}^0
 - \frac{e}{2 m} \svec{\sigma} \cdot \svec{B}_e.
\ee
When spin transition from down to up in  gravity, it will radiate
the following energy
\be \label{2.6}
\Delta E =  \left | g \svec{B}^0  + \frac{e}{m} \svec{B}_e \right |.
\ee
\\

The above results are only applicable to point-like particles.
It is known that neutron and proton are not point-like particle, for
they have inner structures. Now, we extend the above results
to general cases. If we denote the magneton $\svec{\mu}$
of a particle as
\be \label{2.7}
\svec{\mu} = g_L \frac{e}{2 m} \svec{J},
\ee
where $\svec{J}$ is the spin of the particle and $g_L$ is the Lande
$g$-factor, the coupling energy $\Delta H$ should be
\be \label{2.8}
\Delta H =
 - g_L \frac{g}{2} \svec{\sigma} \cdot \svec{B}^0
 - \svec{\mu} \cdot \svec{B}_e.
\ee
Then, the particle will emit the following energy during spin
transition
\be \label{2.9}
\Delta E = \frac{g_L}{2}
\left | g \svec{B}^0  + \frac{e}{m} \svec{B}_e \right |.
\ee
The separations of different Landau levels is  integer
multiples of $\Delta E$.
Detecting such kind of radiation can directly measure the
gravitomagnetic field on the surface of the star.
\\

\section{Determination of Gravitomagnetic Field from GRBs or X-ray Pulsars}
\setcounter{equation}{0}

It is known that, in Gamma-Ray Bursts(GRG), some absorption
spectral lines are detected\cite{18,19,20}. Spectral features
due to electron cyclotron resonance have been discovered
from some X-ray pulsars\cite{21,22,23,24}.
These spectral lines
are traditionally interpreted as cyclotron harmonic in strong
magnetic fields of about $10^{12}$ gauss. Now, according to
equation (\ref{2.6}), gravitomagnetic fields also contribute
to such kind of excitation. It is known that ordinary neutron
stars are spinning with high angular velocity, and their
angular momenta are also very high, so the gravitomagnetic
fields generated by neutron stars should also be quite
large. For a spinning black hole, its gravitomagnetic
field should be very strong near the event horizon. Therefore,
in these cases, the positions of the spectral lines contains
the information of gravitomagnetic fields in the surface of
the star, and we can determine the gravitomagnetic field
through precise measuring the positions of those spectral lines.
But just from one serial of spectral
lines, we can not simultaneous determine the gravitomagnetic
field $|g \svec{B}^0|$ and magnetic field $|\svec{B}_e|$.
So, besides detecting the spectral lines of electrons,
we need also detect the corresponding spectral lines
of protons or neutrons. The masses of proton, neutron
and electron are different, and their energy levels of spin
transitions are also
different. For electron, the position of the lowest spectral
line is
\be \label{3.1}
\Delta E_e = | g_{Le} | \cdot
\left | g \svec{B}^0  + \frac{e}{m_e} \svec{B}_e \right |,
\ee
and those for proton  and neutron are
\be \label{3.2}
\Delta E_p =  | g_{Lp} | \cdot
\left | g \svec{B}^0  + \frac{e}{m_p} \svec{B}_e \right |
\ee
and
\be \label{3.2a}
\Delta E_n =  | g_{Ln} | \cdot
\left | g \svec{B}^0  + \frac{e}{m_n} \svec{B}_e \right |
\ee
respectively,
where $m_e$ $m_p$ and $m_n$ are masses of electron, proton
and neutron respectively,
and $g_{Le}$, $g_{Lp}$ and $g_{Ln}$ are Lande $g$-factors of
electron, proton and neutron respectively. It is know that
\be \label{3.3}
m_e : {m_p} : {m_n} \simeq 1 : 1836 : 1839,
\ee
\be \label{3.3a}
g_{Le} : {g_{Lp}} : {g_{Ln}} \simeq 1  : 2.79 : -1.91.
\ee

If the excitation is pure electromagnetic or dominantly
electromagnetic
\be \label{3.3b}
| g \svec{B}^0  | \ll | \frac{e}{m_e} \svec{B}_e  |,
\ee
then we should observe three serials of spectral lines,
and the positions of the lowest spectral lines of electron,
proton and neutron should obey the ratio
\be \label{3.8}
{\Delta E_e} : {\Delta E_p} : \Delta E_n
\simeq 961 : 1.46 : 1.
\ee
Suppose that the detected spectral lines reported
in literature \cite{18,19} are from excitation
of electrons in strong electromagnetic field,
that is $\Delta E_e$ is about 20 KeV, then
$\Delta E_p$ should be about 30 eV. In this case, we should
also observe some spectral lines with equally spaced at about 30 eV,
60 eV,  $\cdots$. But if those spectral lines are
from excitation of protons, that is $\Delta E_p$ is about 20 KeV,
then $\Delta E_e$ should be about 13 MeV. If we
precisely measure the position of spectral lines and find
that they obeys the above ratio, we can conclude that the
GRBs comes from a neutron star or that X-ray pulsar is not
a black hole, and the gravitomagnetic
field in that neutron star is relatively very weak.
\\

It is known that, at the event horizon, the gravitational gauge
field is divergent, and the field strength of gravitational gauge
field is also divergent. It means that, near the event horizon,
the gravitomagnetic field can be very strong, or can be much
stronger than electromagnetic field. So near a black hole,
the excitation can be pure gravitational or dominantly gravitational, then
\be \label{3.5}
{\Delta E_e} : \Delta E_n : {\Delta E_p} \simeq 1 : 1.91 : 2.79,
\ee
and we should not observe any absorption lines at lower energy region.
In this case, the position of the lowest spectral line of neutron
is close to that of the second harmonic of electron, and the position
of the lowest spectral line of proton is close to the third
harmonic of electron. If the position of the first, the second
and the third spectral lines satisfies the above relation, not
exact $ 1 : 2 : 3$, and there are no harmonic spectral lines at
much lower or much higher positions,
it is possible that the GRB comes from
a black hole or the X-ray pulsar is a black hole.
\\

If both gravitomagnetic field and electromagnetic field contribute
to the excitation, then
\be \label{3.6}
0.36 < \frac{\Delta E_e}{\Delta E_p} < 658.
\ee
Therefor, directly observing the absorption spectral lines
of proton and precise measuring the position of the spectral
line will help us to understand the mechanism of the
excitation and the nature of GRBs or X-ray pulsars. \\

For a spinning black hole, its gravitomagnetic field will
be very strong and its electromagnetic field will be relatively
weak. So, the spin transition or cyclotron harmonic is
dominantly gravitational, and
\be \label{3.7}
\frac{\Delta E_e}{\Delta E_p} \sim 0.36.
\ee
So, the ratio $\frac{\Delta E_e}{\Delta E_p}$ can be served as
a quantitative criteria of black hole. In other words, using
this ratio, we can discriminate black hole from neutron star.
\\

\section{Summary and Discussions}

In gauge theory of gravity, there is direct coupling between
spin and gravitomagnetic field. In GRBs or X-ray pulsars,
absorption spectral
lines of electron were observed. If these lines originate
from strong coupling between magneton and electromagnetic
fields, there must exists new serials of spectral lines
at much lower or much higher positions.
Simultaneously determination of the absorption lines of
electron, proton and neutron can help us to determine both the
gravitomagnetic field and electromagnetic field, and help
us to understand the nature of GRBs or X-ray pulsars.
The ratio of the position of the lowest spectral line
can be served as a criteria of black hole. \\

In order to observe the absorption spectral lines of both
electron and proton, experimental observation should be
performed in a large scope of energy region, at least from
1 eV to 100 MeV, which covers the spectral range from
visible light to $\gamma$ rays. Therefore, a combine
observation of astronomical telescope, X-ray telescope
and $\gamma$-ray telescope  is needed.
\\

\end{document}